\documentclass{article}
\usepackage{latexsym}
\usepackage{graphicx}
\usepackage{mathptmx}

\usepackage{amsmath}
\usepackage{amsfonts}
\usepackage{amssymb}
\usepackage{amsbsy}
\usepackage{amsthm}
\usepackage{multirow}

\usepackage{listings}
\usepackage{color}

\definecolor{dkgreen}{rgb}{0,0.6,0}
\definecolor{gray}{rgb}{0.5,0.5,0.5}
\definecolor{mauve}{rgb}{0.58,0,0.82}

\usepackage[pdftex,colorlinks=true,urlcolor=blue,citecolor=black,anchorcolor=black,linkcolor=black]{hyperref}

\renewcommand{\emph}[1]{{\bf#1}}
\newcommand{\FX}{\bar{F}_X^{-1}}
\newcommand{\FY}{\bar{F}_Y^{-1}}
\newcommand{\Var}{\mathop{\text{Var}}}
\newcommand{\E}{\mathop{\text{E}}}
\newcommand{\R}{{\mathbb R}}
\usepackage{fullpage}

\newtheoremstyle{wsc}
{3pt}
{3pt}
{}
{}
{\bf}
{}
{.5em}
{}

\theoremstyle{wsc}

\begin{document}

%
%

\title{IMPORTANCE SAMPLING FOR THE SIMULATION OF REINSURANCE LOSSES}

\author{Georg Hofmann
\\ [12pt]
Validus Research Inc.\\
Suite 201, 187 King Street South\\
Waterloo Ontario, N2J 1R1, Canada}


\maketitle


\section*{ABSTRACT}
      Importance sampling is a well developed method in statistics. Given a random variable $X$, the problem of estimating its expected value $\mu$ is addressed. The standard approach is to use the sample mean as an estimator $\overline x$. In importance sampling, a suitable variable $L$ is introduced such that the random variable $X/L$ has an estimator with a smaller variance than that of $\overline x$. As a result, a smaller sample size can lead to the same estimation accuracy. 
      
      In the simulation of reinsurance financial terms for catastrophe loss, choosing a general variable $L$ is difficult: Even before the application of financial terms, the loss distribution is often not modelled by a closed-form distribution. After that, a wide range of financial terms can be applied that makes the final distribution unpredictable. However, it is evident that the heavy tail of the resulting net loss distribution makes the use of importance sampling desirable. We propose an importance sampling technique using a power function transformation on the cumulative distribution function. The benefit of this technique is that no prior knowledge of the loss distribution is required. It is a new technique that has not been documented in the literature. The transformation depends on the choice of the exponent $k$. For a specific example we investigate desirable values of $k$.
      
\section{INTRODUCTION}
\label{sec:intro}

    In many applications heavy-tailed distributions are sampled according to the Monte Carlo method in order to approximate the mean (or other metrics). The accuracy of such simulations can be improved by increasing the sample size. However, in practice this is often expensive in terms of performance and computer resources. This is due to the fact that the simulation error decreases only very slowly as the sample size increases. 
    
    Let $X$ be a random variable whose variance $\Var(X)$ exists. Its expectation $E(X)$ can be estimated by taking the mean of a sample with sample size $n$. The standard error of this estimator $\overline{X}_n$ is given by
    \begin{align*}
    \sqrt{\Var(\overline{X}_n)}&=\frac{\sqrt{\Var(X)}}{\sqrt{n}}.
    \end{align*}
    So to reduce the standard error by a factor of $m$ (and thus increase accuracy) the sample size has to be multiplied by $m^2$.
    
    Another way to improve accuracy is to reduce the variance $\Var(X)$. The standard approach for this in the literature is \emph{importance sampling}. See for example \cite{bucklew2004introduction} for a recent textbook on this subject. For this approach a random variable $L\ge0$ is chosen such that $\E(L)=1$. If $P$ is the given probability measure, a new one can be defined as $P^{(L)}=LP$. With respect to the two probability measures we have the following equality in expectation:
    \begin{align*}
    \E(X;P)&= \E\left(\frac{X}{L};P^{(L)}\right).
    \end{align*}
    If $L$ is chosen suitably, then we obtain a reduction in variance:
    \begin{align*}
    \Var(X;P)&> \Var\left(\frac{X}{L};P^{(L)}\right).
    \end{align*}

    In this article we propose an application of importance sampling to the evaluation of catastrophe loss risk for reinsurance companies. This evaluation is usually accomplished in roughly two steps using commercial catastrophe models: First the \emph{gross} loss is determined for a stochastic ensemble of event occurrences. Then this gross loss is subjected to reinsurance financial terms of a reinsurance contract. This happens during an annual simulation that may have several events occur in a simulated year. The final result of this process are metrics describing the annual \emph{net} loss distribution for the reinsurance contract. 
    
    The challenge here is that it is impossible to determine a single distribution for $L$ that provides effective importance sampling for all kinds of catastrophe risk and reinsurance contracts. This is due to the fact that the distribution of the net loss is not given in a closed form. It depends heavily on the science used in the catastrophe model and on the type of reinsurance contracts considered. Nonetheless it is clear that the distribution has a significant tail that promises a significant gain from importance sampling.
    
Our approach is to use a simple form of importance sampling on the loss severity resulting from the event ensemble. The uncertainty around this loss is generally referred to in the insurance industry as \emph{rate uncertainty} or \emph{primary uncertainty} \cite{CorSimUncRMS}. Going forward, we will denote this loss severity by $X$. However the effectiveness of the importance sampling will be measured by the simulation error associated with metrics of the annual loss.

    If $F_X$ is the cumulative distribution function of $X$, then its quantile function $F_X^{-1}:[0,1]\to\R$ is defined by
        \begin{equation*}
          F_X^{-1}(p)~=~\inf\{x\in\R:p\le F(x)\}.
        \end{equation*}
    See \cite{gilchrist2000statistical} for a recent textbook that advocates the usage of quantile functions for statistical modelling. The expected value of $X$ can be calculated as follows:
    \begin{eqnarray}\label{eq:IntroEX}
      \E(X) 
      &=& \int_0^1F_X^{-1}(p)~dp
    \end{eqnarray}
    For a given sample size $n$ let $p_1$, $p_2,\dots,p_n$ be samples from the equidistant partition. Then we have the following Riemann sum approximation:
    \begin{align*}
      \E(X)
      &\approx
      \frac1n\sum_{i=1}^nF_X^{-1}(p_i)
    \end{align*}
    This approximation may also be thought of as the Monte Carlo method if $p_1$, $p_2,\dots,p_n$ constitute a random uniform sample. We will achieve the reduction in variance discussed earlier by performing a substitution in the variable $p$ of Equation~(\ref{eq:IntroEX}).
    
    This article is intended to be useful for professionals with a working knowledge of random simulations. The necessary formulae for our proposed sampling method and its error estimation can be found in Subsection~\ref{sec:RiemannSums}. The relevant equations are (\ref{eq:RSEXSub}) and (\ref{eq:RSEXSquared}). In Section~\ref{sec:Example} we have included a case study to measure the effectiveness of the proposed importance sampling approach. The R code used to evaluate this case study can be found in the appendix. 

    \section{IMPORTANCE SAMPLING USING A POWER FUNCTION SUBSTITUTION}
    In this section we work out the the mathematical details necessary to specify the proposed importance sampling approach. Subsection~\ref{sec:ASubstitution} contains the theoretical basis for our approach. Subsection~\ref{sec:RiemannSums} provides the details about the application of the approach using Riemann sampling. Subsection~\ref{sec:MonteCarlo} explains how to transition from Riemann sampling to random sampling and explains error estimates. 

    \subsection{A Substitution}\label{sec:ASubstitution}
    \label{sec:ASubstitution}

    In this subsection, we introduce a family of transformations that will lead to a variance reduction in some cases. It is therefore our goal to compare the original variance $\Var(X)$ with the variance $\Var(Y)$ in the transformed case. Note that all random variables are real-valued. 
    
    Let $X$ be a random variable for which the expected value $\E(X)$ and the variance $\Var(X)$ exist. We denote its cumulative distribution function by $F_X$. Then its tail distribution is defined as 
    \begin{align*}
    \bar{F}_X(x)&=1-F_X(x).
    \end{align*}
    We will work with $\bar{F}(x)$ rather than $F_X$ for the following numerical reason: The accuracy of floating point numbers on computers is higher in the vicinity of 0 than in the vicinity of 1. Since $\bar{F}_X$ encodes information about the tail of the distribution in values around 0, it is preferable to $F_X$.
    
    The complementary quantile function is the function $\FX:[0,1]\to\R$ defined by
    \begin{equation*}
      \FX(p)~=~\inf\{x\in\R:p\le\bar{F}(x)\}.
    \end{equation*}
    Now let 
    \begin{equation*}
      t: [0,1]\to[0,1]
    \end{equation*}
    be a differentiable, strictly increasing function on the unit interval that satisfies $t(0)=0$ and $t(1)=1$. Later we will specialize $t$ to the function defined by $t(q)=q^2$, so you may keep this example in mind.

    Now let $Y$ be a random variable with the complementary quantile function defined by
    \begin{equation*}
      \FY(q)~=~\FX(t(q))t'(q).
    \end{equation*}
    The random variable $Y$ could be realized, for instance, by taking a random variable $U$ that is uniformly distributed on the unit interval. Then set $Y=\FY(U)$.
    
    By using the substitution $p=t(q)$ we obtain
    \begin{eqnarray}
      \E(X) 
      &=& \int_0^1\FX(p)~dp\label{eq:EX} \\
      &=& \int_0^1\underbrace{\FX(t(q))t'(q)}_{=\FY(q)}~dq\label{eq:EXSub} 
      ~=~ \E(Y)
    \end{eqnarray}
    So, the two random variables $X$ and $Y$ have the same expected value. We will compute the difference in their variances. To accomplish this we first compute
    
    \begin{eqnarray}
      \E(X^2) 
      &=& \int_0^1\big(\FX(p)\big)^2~dp\nonumber\\
      &=&\int_0^1\Big(\FX\big(t(q)\big)\Big)^2t'(q)~dq\label{eq:EXSquared}\\
      \E(Y^2) 
      &=&\int_0^1\Big(\FX\big(t(q)\big)\Big)^2\big(t'(q)\big)^2~dq\label{eq:EYSquared}\\
      &=& \int_0^1\big(\FX(p)\big)^2\frac{\Big(t'\big(t^{-1}(p)\big)\Big)^2}{t'\big(t^{-1}(p)\big)}~dp\nonumber\\
      &=&\int_0^1\big(\FX(p)\big)^2t'\big(t^{-1}(p)\big)~dp.
    \end{eqnarray}
    Since $\Var(x)=\E(X^2)-\big(\E(X))^2$ and $\E(X)=\E(Y)$, we obtain
    \begin{eqnarray*}
       \Var(Y)-\Var(X)
      &=& \E(Y^2)-\E(X^2)\\
      &=& \int_0^1\big(\FX(p)\big)^2\Big(t'\big(t^{-1}(p)\big)-1\Big)~dp.
    \end{eqnarray*}
    Our objective is to achieve a reduction in variance. This happens when the above expression is negative. To further understand when this occurs, it is instructive to investigate the term $\Big(t'\big(t^{-1}(p)\big)-1\Big)$ for its sign. 
    
    Let $k\in[1,\infty)$. Our primary example of a selection of $t$ is defined by
    \begin{align}\label{eq:specialT}
    t(q)&=q^k.
    \end{align}
    We compute
    \begin{eqnarray*}
    t'(q)&=&kq^{k-1},\\
    t^{-1}(p)&=&p^{\frac1k},\\
    t'\big(t^{-1}(p)\big)&=&kp^{\frac{k-1}k}.
    \end{eqnarray*}
    For the specific case $k=2$, the last term is $2\sqrt{p}$. In this case a reduction of variance will be achieved if the following inequality of positive integrals is satisfied:
    \begin{align}
            \int_0^{1/4}\big(\FX(p)\big)^2(1-2\sqrt{p})~dp
        ~>~
        \int_{1/4}^1\big(\FX(p)\big)^2(2\sqrt{p}-1)~dp.
    \end{align}
    This is the case, if $\big(\FX(p)\big)^2$ is large enough on the interval $[0,1/4]$. That, in turn, happens if the distribution of $X$ has significant enough a tail.
    

    \subsection{Riemann Sums}
    \label{sec:RiemannSums}
    Let $n$ be the sample size and $(r_i)_{i=1}^n$ the equidistant midpoint sample points given by
    \begin{eqnarray*}
      r_i&=& \frac{i-\frac12}{n}
      ~~~\text{for}~ i=1,2,3,\dots,n.
    \end{eqnarray*}
    The expected value $\E(X)$ can be approximated by the middle Riemann sum. This can be done using either (\ref{eq:EX}):
    \begin{eqnarray}
      \E(X)
      &\approx&
      \frac1{n}\sum_{i=1}^n\FX(r_i),\label{eq:RSEX}
    \end{eqnarray}
    or it can be done using (\ref{eq:EXSub}):
    \begin{eqnarray*}
      \E(X)      
      &\approx&
      \frac1{n}\sum_{i=1}^n\FX(t(r_i))t'(r_i).
    \end{eqnarray*}
    If we define the \emph{sample values} $s_i=\FX(t(r_i))$  and the \emph{sample weights} $w_i=t'(r_i)$ then the above equation takes the simple form
    \begin{eqnarray}
      \E(X)
      &\approx& 
      \frac1n\sum_{i=1}^n s_iw_i. \label{eq:RSEXSub}
    \end{eqnarray}
    In the same way, we can compute middle Riemann sums for the following quantities. We use Equations~(\ref{eq:EXSquared}) and (\ref{eq:EYSquared}) for this:
    \begin{eqnarray}
      \E(X^2)
      &\approx& 
      \frac1n\sum_{i=1}^n s_i^2w_i,\label{eq:RSEXSquared}\\
      \E(Y^2)
      &\approx& 
      \frac1n\sum_{i=1}^n (s_iw_i)^2.\label{eq:RSEYSquared}
    \end{eqnarray}

    We define the \emph{sample improvement} by
    
    \begin{eqnarray*}
      I
      ~=~
      \frac{\Var(X)}{\Var(Y)}
      &=&
      \frac{\E(X^2)-\big(\E(X)\big)^2}{\E(Y^2)-\big(\E(X)\big)^2}.
    \end{eqnarray*}

    \subsection{Monte Carlo Method}
    \label{sec:MonteCarlo}

    The formulae (\ref{eq:RSEX}) to (\ref{eq:RSEYSquared}) can be used as a Monte Carlo method to approximate the expected value of $X$ and the variances of $X$ and $Y$. In that case  $(r_i)_{i=1}^n$ is uniformly distributed between 0 and 1. Either methods (\ref{eq:RSEX}) and (\ref{eq:RSEXSub}) can be used to approximate $E(X)$. We will refer to the two methods as the \emph{regular method} and the \emph{enhanced method}. If the transformation $t$ is chosen appropriately for the distribution of $X$ then the sample improvement $I$ is greater than 1 and the variance of $Y$ is less than the variance of $X$. The error estimates are given as follows:
    \begin{eqnarray*}
    \text{Regular method:}&&\sqrt{\frac{\Var(X)}{n}}\\
    \text{Enhanced method:}&&\sqrt{\frac{\Var(Y)}{n}}
    \end{eqnarray*}
    One way to interpret the sample improvement $I$ is that it is the number by which you would need to multiply the sample size for the regular method to catch up to the enhanced method. 

    \section{AN EXAMPLE OF CATASTROPHE REINSURANCE RISK}
    \label{sec:Example}
    In this section we use a model to quantify the distribution of gross property insurance losses resulting from catastrophes. Then we apply a typical set of reinsurance contracts to arrive at a net loss distribution. This model is not based on catastrophe science, so it can not be used to base any business decisions on. However, the model is realistic enough to measure the effectiveness of the proposed importance sampling approach. From there the reader may draw conclusions about what to expect from this approach for more advanced models with similar distributions. The data presented in Table~\ref{tab:Contracts} and Figure~\ref{fig:SampleImprovement} was obtained using the R code in the appendix. A trial number of 100 million was used for that, however the simulation error was expressed as an error for a 1 million trial simulation.
	
    The simple model that we use here is based on the following assumptions:
    \par\bigskip
	\begin{center}
	\begin{tabular}{|l|r|}
	\hline
	  \begin{minipage}[t]{1.8cm}
	  	  Event \\ Occurrence
	  \end{minipage} 
	  &
	  \begin{minipage}[t]{8cm}
	  The occurrence of events throughout a year is modelled according to a Poisson distribution with frequency $\lambda=3$.
	  \smallskip
	  \end{minipage}
	  \\
	\hline
	  \begin{minipage}[t]{1.8cm}
	  	  Gross Loss \\ Severity
	  \end{minipage} 
	  &
	  \begin{minipage}[t]{8cm}
	  The gross loss of an event is modelled according to a log-normal distribution fitted to a mean of 10 and a standard deviation of 30. 
	  \smallskip
	   
	  \end{minipage}
	  \\
    \hline
	\end{tabular}
	\end{center}

	\bigskip
	It is important to note that for larger values of $\lambda$ the importance sampling approach will be subject to what is known as the \emph{Curse of Dimensionality} \cite{Li_curse-of-dimensionalityrevisited}. In that case the stability of the simulation decreases as $k$ increases.

    \begin{table*}[t]
    
      \centering
      \bigskip
      \begin{tabular}{l|rrrrrrrrrr}
      \cline{2-11}
      &\multicolumn{2}{|p{1.6cm}|}{\centering Occurrence Terms}
      &\multicolumn{2}{|p{1.6cm}|}{\centering Annual Aggregate Terms}
      &\multicolumn{2}{|p{2cm}|}{\centering Simulation Result}
      &\multicolumn{4}{|p{5.1cm}|}{\centering Simulation Error (Radius of 95\% Confidence Interval)}
      \\\cline{2-11}
      &\multicolumn{1}{|p{0.8cm}|}{\centering Att}
      &\multicolumn{1}{|p{0.8cm}|}{\centering Lim}
      &\multicolumn{1}{|p{0.8cm}|}{\centering Att}
      &\multicolumn{1}{|p{0.8cm}|}{\centering Lim}
      &\multicolumn{1}{|p{1cm}|}{\centering EL}
      &\multicolumn{1}{|p{1cm}|}{\centering EL\%}
      &\multicolumn{1}{|p{1cm}|}{\centering $k=1$}
      &\multicolumn{1}{|p{1.1cm}|}{\centering $k=1.5$}
      &\multicolumn{1}{|p{1cm}|}{\centering $k=2$}
      &\multicolumn{1}{|p{1cm}|}{\centering $k=3$}
      \\\hline
    Contract 1&34&34&0&34&3.457&10.17\%&0.54\%&0.37\%&0.42\%&0.74\%\\
    Contract 2&95&95&0&95&1.879&1.98\%&1.25\%&0.65\%&0.63\%&0.92\%\\
    Contract 3&190&190&0&190&0.969&0.51\%&2.44\%&1.03\%&0.88\%&1.13\%\\
    Contract 4&9&9&9&9&0.865&9.61\%&0.56\%&0.42\%&0.52\%&1.03\%\\
    Contract 5&20&20&20&20&0.391&1.96\%&1.26\%&0.68\%&0.70\%&1.14\%\\
    Contract 6&34&34&34&34&0.168&0.49\%&2.46\%&1.03\%&0.93\%&1.37\%\\

      \end{tabular}
      \bigskip
      \caption{Evaluation of reinsurance contracts based on a 1 million trial simulation.}
      \label{tab:Contracts}
    \end{table*}
	
	Table~\ref{tab:Contracts} shows the details of 6 typical reinsurance contracts. Occurrence attachments and limits are applied to the gross loss after every event occurrence. The resulting loss is accumulated within every trial year and then subjected to the annual aggregate attachments and limits. This results in the net loss. The expected loss (EL) of a contract is often expressed as a percentage of the occurrence limit. We denote this percentage by EL\%. According to their definitions the first three contracts are first-event covers and the last three are second-event covers.  In each category there are contracts that have an approximate EL\% of 0.5\%, 2\% and 10\%, which are typical numbers for the reinsurance industry. The more volatile contracts are those with a lower EL\%. As a consequence, they also show the largest simulation error.
	  
Figure~\ref{fig:SampleImprovement} shows that the best sample improvements are obtained for values of $k$ between 1.5 and 2. But if the objective is to keep the simulation error of any contract below a certain threshold, then $k=2$ is a good choice, as it addresses the simulation errors of the more volatile contracts 3 and 6 in the best possible way. This can also be seen in Table~\ref{tab:Contracts}, where the choice of $k=2$ reduces all simulation errors below 1\%.

    \section{CONCLUSION}
    Important metrics in risk management are often approximated using random or Riemann sampling. In this article we present an importance sampling approach that allows an increase in the accuracy of such approximations. It can be used to reduce the error margin of results or to reduce the run time while maintaining the same accuracy. The sample improvement $I$ can be viewed as the factor by which the run time can be reduced. For a typical catastrophic risk and some typical reinsurance contracts the sample improvement is calculated. Based on this analysis , the value $k=2$ of the tuning parameter $k$ is recommended. However, the choice of $k$ may vary with the distribution at hand. A simple modification of the included code would allow an investigation of other distributions for values of $k$ with maximal gain in performance.
    \begin{figure*}[t]
      \centering
      \par\includegraphics[width=16cm]{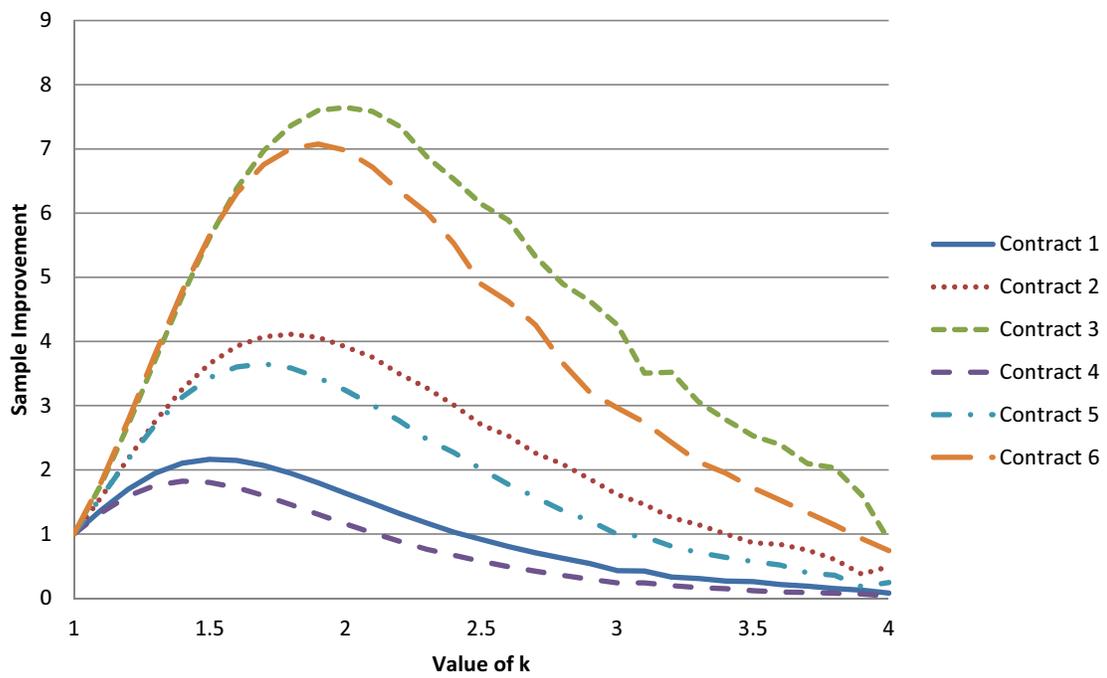}  
      \par\vspace{-1.5cm}   
      \caption{Sample improvement for 6 different contracts.}
      \label{fig:SampleImprovement}
    \end{figure*}

    \appendix
    \onecolumn

    \section{R CODE}

    The following code runs with R-2.13.1. No additional packages need to be installed.

      \lstset{language=R}
      \begin{lstlisting}
# This code belongs to the article 
# "IMPORTANCE SAMPLING FOR THE SIMULATION OF REINSURANCE LOSSES".

### Initial Parameters

loss.sevty.mean <- 10
loss.sd <- 30
total.freq <- 3.0
k.sevty.vector <- (10:40) / 10
confidence.level <- .95
num.trials <- 1E5
random.sampling <- FALSE #TRUE = random sampling, FALSE = Riemann sampling
occ.attach <- c(   34,   95,  190,   9,   20,   34) 
occ.lim    <- c(   34,   95,  190,   9,   20,   34)
agg.attach <- c(    0,    0,    0,   9,   20,   34)
agg.lim    <- c(   34,   95,  190,   9,   20,   34)

### Main Code

# Fit a log-normal distribution with the desired mean and standard deviation:
loss.sigma <- sqrt(log((loss.sd / loss.sevty.mean) ^ 2 + 1))
loss.mu <- log(loss.sevty.mean) - loss.sigma ^ 2 / 2
# Calculate confidence radius using a normal distribution assumption:
confidence.radius <- qnorm(p=1 - (1 - confidence.level) / 2, mean=0, sd=1)
# Equidistant partition and midpoint sample:
freq.riemann.sample <- ((1:num.trials)-.5)/num.trials
# Poison Frequency:
freqs <- qpois(p=freq.riemann.sample, lambda=total.freq, lower.tail=FALSE)
num.events <- sum(freqs)
# Initialize table of results
result.table <- data.frame(occ.attach, occ.lim, agg.attach, agg.lim)
# Loop over provided values of k:
for(k.counter in 1:length(k.sevty.vector)){
  k.sevty <- k.sevty.vector[k.counter]
  if(random.sampling){
    sevty.random.sample <- runif(n=num.events)
    event.loss <- 
      qlnorm(p=sevty.random.sample ^ k.sevty, meanlog=loss.mu, 
             sdlog=loss.sigma, lower.tail=FALSE)
    sevty.weights <- k.sevty * sevty.random.sample ^ (k.sevty - 1)
  }else{
    sevty.riemann.sample <- ((1:num.events)-.5)/num.events
    event.loss <- 
      qlnorm(p=sevty.riemann.sample ^ k.sevty, meanlog=loss.mu, 
             sdlog=loss.sigma, lower.tail=FALSE)
    sevty.weights <- k.sevty * sevty.riemann.sample ^ (k.sevty - 1)
    # Reorder the event occurences randomly:
    reordering.sample <- sample(x=num.events)
    event.loss <- event.loss[reordering.sample]
    sevty.weights <- sevty.weights[reordering.sample]
  }
  last.event.of.trial <- cumsum(freqs)
  trial.weight <- 
    exp(diff(c(0,cumsum(c(0,log(sevty.weights)))[last.event.of.trial + 1L])))
  num.terms <- length(occ.attach)
  for(term.counter in 1:num.terms){
    event.loss.occ.layered <- 
      pmin(pmax(0, event.loss - occ.attach[term.counter]), occ.lim[term.counter])
    event.loss.cumulative <- c(0, cumsum(event.loss.occ.layered))
    trial.loss.cumulative <- c(0 ,event.loss.cumulative[last.event.of.trial+1])
    trial.loss <- diff(trial.loss.cumulative)
    trial.loss.layered <- 
      pmin(pmax(0, trial.loss - agg.attach[term.counter]), agg.lim[term.counter])
    mean.loss <- (trial.loss.layered %*% trial.weight) / num.trials
    mean.loss.squared <- (trial.loss.layered ^ 2 %*% trial.weight) / num.trials
    sd.loss <- sqrt(mean.loss.squared - mean.loss ^ 2)
    sim.error.regular <- confidence.radius * 
      sd.loss / sqrt(num.trials) / mean.loss
    sim.error.enhanced <- confidence.radius * 
      sd(trial.loss.layered * trial.weight) / sqrt(num.trials) / mean.loss
    if(k.counter==1){
      result.table$mean.loss[term.counter] <- mean.loss
      result.table$mean.loss.percent[term.counter] <- 
        100 * mean.loss / occ.lim[term.counter]
      result.table$sim.error.regular[term.counter] <- sim.error.regular
      result.table$sim.error.enhanced[term.counter] <- sim.error.enhanced
    }
    sample.improvement.name <- paste("sample.improvement.k", k.sevty, sep="")
    result.table[term.counter,sample.improvement.name] <- 
      (sim.error.regular / sim.error.enhanced) ^ 2  
  }
}
print(round(result.table,2))
      \end{lstlisting}

    \bibliographystyle{alpha}
    \bibliography{../../../Latex/references/references}

\section*{AUTHOR BIOGRAPHY}

\noindent {\bf GEORG HOFMANN} is a member of the research team at Validus Research Inc., Waterloo Ontario, Canada. He is an adjunct professor at the Department of Mathematics and Statistics of Dalhousie University, Nova Scotia, Canada and a technical advisor of the Risk Analytics Lab of the Computer Science Department. He has worked in the reinsurance industry for 5 years as a data scientist, gaining experience in the science and implementation of catastrophe risk models. He received a Mathematics Diploma (Masters) with a Minor in Quantum Mechanics from the Technische Universit\"at Darmstadt (Germany). At the same university he received his Ph.D. in Mathematics.

\end{document}